**A study on the mechanical property and corrosion sensitivity of an AA5086 friction stir welded joint**

Zhitong Chen*, Shengxi Li, Kaimiao Liu, Lloyd H. Hihara*

Hawaii Corrosion Laboratory, Department of Mechanical Engineering, University of Hawaii at Manoa, Honolulu, HI 96822, USA

*Corresponding author: jeadonchen@gmail.com (Z. Chen); hihara@hawaii.edu (L.H. Hihara)

**Abstract**

The mechanical property and corrosion sensitivity of a friction stir welded (FSW) AA5086 joint have been investigated. Vickers microhardness and tensile tests showed that the TMAZ/HAZ zone had the lowest microhardness value and the tensile specimen (transverse to the weld) failed in this region. Polarization experiments conducted using samples cut from each zone indicated that the HAZ had the lowest corrosion rate in 3.15 wt.% NaCl. SEM observation of the samples, Raman spectroscopic and XRD characterization of the corrosion products, and weight loss analysis after immersion tests all showed that the corrosion severity of AA5086 decreased from 3.15% NaCl to ASTM seawater and finally to 0.5 M $Na_2SO_4$. In addition, less corrosion products were observed from HAZ as compared to the other three regions, which was in agreement with the Raman and XRD analyses.

# 1. Introduction

High strength aluminum alloys such as the 5xxx series are widely used in automotive, aerospace and ship industries because of their light weight and high strength. These alloys are difficult to weld using conventional techniques such as fusion-based welding. The development of friction stir welding (FSW) - a solid state joining techniques - has provided an improved way of joining aluminum alloys. Since its invention by The Welding Institute (TWI) of UK in 1991 [1], FSW has been considered as one of the most significant metal joining technologies. Generally, FSW improves mechanical properties of the joints and creates minimal microstructural changes at the weld, as compared to conventional joining techniques including tungsten inert gas (TIG) and metal inert gas (MIG) welding [2].

The FSW joint of aluminum alloys usually contain four distinct microstructural zones: the Nugget zone (NZ), the heat-affected zone (HAZ), the thermomechanically-affected zone (TMAZ), and the base material (BM) [3-8]. The NZ is the region where the tool piece pin passes, and thus experiences high deformation and high heat. It has fine equiaxed grains due to full recrystallization. The HAZ experiences only a heating effect, without any mechanical deformation. The TMAZ adjacent to the NZ is the region where the metal is plastically deformed and heated, which was, however, not sufficient to cause recrystallization. The microstructural changes induced by the plastic deformation and the frictional heat of FSW process can have detrimental effects on the thermomechanical, corrosion, and fatigue properties of the weld-affected regions [9].

The microstructure and mechanical properties of FSW joints of 5xxx series aluminum alloys have been extensively investigated [6, 9-17]. Peel et al. studied microstructural, mechanical properties and residual stresses as a function of welding speed in AA5083 friction stir welds [15]. They claimed that recrystallization resulted in the weld zone having considerably lower hardness and yield stress than the parent AA5083. During tensile testing, almost all the plastic flow occurred within the recrystallized weld zone. Cam et al. investigated mechanical properties of friction stir butt-welded AA5086-H32 plate [10, 11]. They found that the heat-induced softening of cold-worked material resulted in lower strength performance, i.e. 75% of the value for the base material. The relatively low ductility was possibly caused by a combination of kissing-bond type defect in the joint and loss of strength within the stirred zone, resulting in confined plasticity. Ramesh et al. focused on multipass friction-stir processing and its effect on mechanical properties of AA5086 [6]. They showed that the hardness of the processed material was higher than the base material. The yield strength and ultimate tensile strength values are lower in multipass-processed material than the base material and single pass-processed materials for the processing conducted at a lower traverse speed. Despite extensive research on the mechanical properties of FSW joints of 5xxx series aluminum alloys, little is known about the corrosion performance of such FSW joints. In the present work, the corrosion behavior of FSW joints of AA5086 was investigated by polarization measurements and immersion tests. Mechanical measurements were also carried out.

## 2. Experimental

AA5086-H32 (0.4% Si, 0.5% Fe, 0.1% Cu, 0.35% Mn, 4.0% Mg, 0.15% Cr, 0.25% Zn, 0.15% Ti, balance Al) plates (300 mm × 50 mm × 6 mm) were friction stir welded vertical to the rolling direction. The traveling speed and rotation speed of the FSW tool were 20 mm/min and 1000 rpm, respectively. The tool had a shoulder diameter of 25 mm. The friction stir pin had a diameter of 8 mm and a height of 6.35 mm. A simultaneous rotation and translation motion of the FSW tool generates the formation of an asymmetric weld. When the tool rotates in the direction of its translation, it refers to the advancing side (AS). When rotation and translation of the tool are in the opposite direction, it refers to the retreating side (RS).

Vickers microhardness testing (Wilson Rockwell, R5000) was performed on the top surface across the FSW weld at a distance of 2 mm using a 50 g load. The tensile tests were performed at a crosshead speed of 3 mm/min using an Instron-5500R testing machine. Tensile specimens were machined from the weld in two directions: longitudinal and transverse (Fig. 1). An axial extensometer with 25 mm gage length was attached to the specimens at the gauge section. The strain analysis of each specimen was made by an ASAME automatic strain measuring system. The tensile properties of the joints were evaluated using three tensile specimens cut from the same joint.

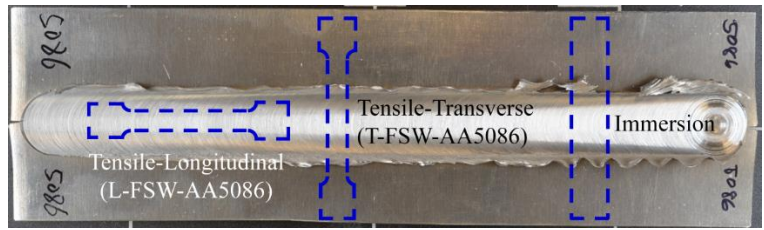

Fig. 1. Schematic diagram showing the locations from where the tensile test (L-FSW-AA5086 and T-FSW-AA5086) and immersion specimens were obtained.

Polarization experiments were conducted using samples (5 mm × 5 mm) cut from different zones, i.e., NZ, TMAZ, HAZ, and BM. The samples were mounted in epoxy resin, polished to a 0.05 μm mirror-finish, and immersed in high purity water (18.0 MΩ·cm) prior to polarization experiments in 3.15 wt.% NaCl solutions at 30 °C. The solutions were deaerated with high-purity nitrogen (> 99.999%). Potentiodynamic polarization experiments were conducted with a PARSTAT 2273 potentiostat (Princeton Applied Research). The working electrodes were kept in the open-circuit condition for 1 hour prior to conducting the potentiodynamic scan at a rate of 1 mV/s. A saturated calomel electrode (SCE) was used as the reference electrode and a platinum mesh was used as the counter electrode. To minimize contamination of the solution, the reference electrode was kept in a separate cell connected via a Luggin probe. Anodic and cathodic sweeps were measured separately on freshly-prepared samples. Polarization experiments for each typical zone were performed using at least three samples to verify reproducibility. To generate polarization diagrams, the mean values of the logarithm of the current density were plotted as a function of potential.

For immersion tests, the FSW samples (70 mm $\times$ 25 mm $\times$ 4 mm) were degreased in acetone, ultrasonically cleaned in deionized water, dried, and weighted prior to the experiments. Three samples from each zone were placed in a 250 ml beaker. A total of 24 beakers were kept at a controlled temperature of 30 ℃ by immersing the beakers in a heated and water-circulated aquarium. Approximately 200 ml of solution (3.15 wt.% NaCl, 0.5 M $Na_2SO_4$, and ASTM seawater) was poured into the beakers so that the samples were completely immersed. Beakers were partially covered to minimize evaporation and to maintain an aerated condition. After 90 and 120 days of immersion, the samples were retrieved and dried in a dry box (1% RH). The corrosion products were analyzed using scanning electron microscopy (SEM, Hitachi S-3400N), Raman spectroscopy (Nicolet Almega XR, Thermo Scientific Corp.), and X-ray diffraction (XRD, Rigaku MiniFlex$^{TM}$). The corroded samples were cleaned in a solution of phosphoric acid ($H_3PO_4$) and chromium trioxide ($CrO_3$) at 90 ℃ for 10 minutes as described in ASTM G01-03. The weight loss of the samples were obtained by recording the weight of the samples before immersion and after chemical cleaning.

## 3. Results and discussion

### 3.1. Weld zone identification

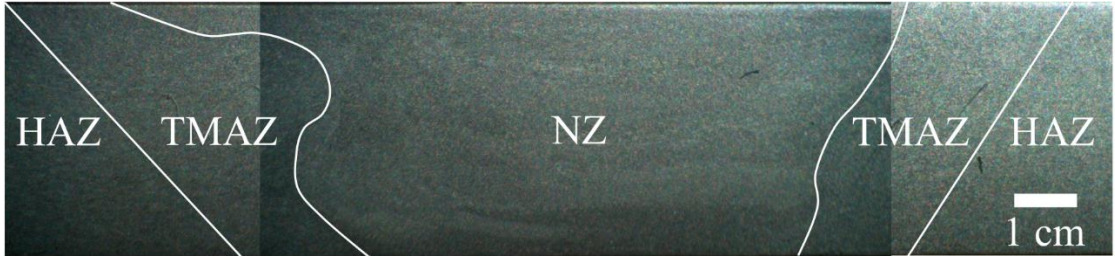

Fig. 2. Cross-sectional view of FSW AA5086 showing typical weld zones (NZ, HAZ, and TMAZ).

Fig. 2 shows the metallographic cross-section of a FSW-AA5086 sample after slight corrosion in acid solution (2.5 vol.% $HNO_3$). The central NZ has microstructure commonly denoted as onion rings. The NZ is wider on the crown region of the weld because the upper surface was in contact with the tool shoulder.

## 3.2. Mechanical properties

Fig. 3 shows the microhardness distribution across the top surface of the FSW-AA5086. The hardness curve demonstrates a "W"-shaped hardness distribution, which is asymmetrical with respect to the weld centerline. The minimum hardness was observed in the TMAZ/HAZ regions, while the maximum value occurred in the BM. According to the Hall-Petch relationship, the hardness value in the NZ was higher than that of the TMAZ/HAZ because of the fine equiaxed grain structure in the NZ [18]. In addition, because AA5086 is strengthened by strain hardening and not by the presence of precipitates, it is possible that the thermal cycle of FSW causes the annihilation of dislocations and destroys the strain hardening mechanism.

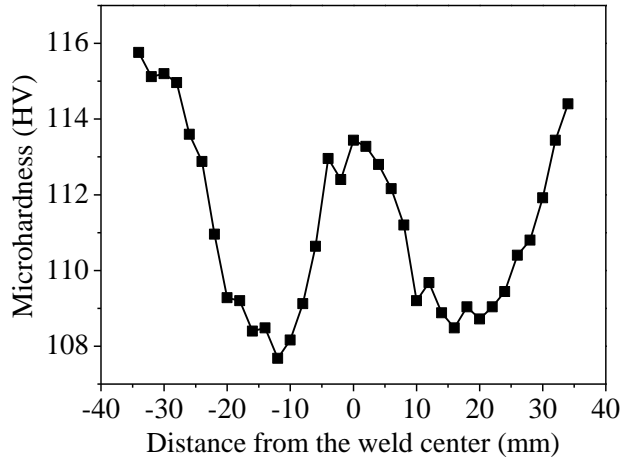

Fig. 3. Microhardness distribution across the top surface of FSW-AA5086 measured with a 2-mm step.

The tensile stress-strain curves of the BM and two types of FSW specimens (Fig. 1) are shown in Fig. 4. The elongation, yield strength, and tensile strength of the BM-AA5086 were approximately 16.7%, 237 MPa, and 300 MPa, respectively. As compared to the BM, all FSW tensile specimens had lower tensile and yield strength values, which can be attributed to the weld zone having lower hardness than the BM. However, the ductility of the two FSW specimens increased as compared to that of the BM. Notice that the longitudinal tensile specimen (L-FSW-AA5086) had a significant increase in ductility with an elongation value more than 2 times of that of the BM. Because the longitudinal tensile specimens contained only fine equiaxed grains from the NZ (Fig. 1), they had better ductility than the BM with relatively large grains. The transverse tensile specimens contained all four zones (i.e., BM, HAZ, TMAZ, and NZ) which have different resistances to deformation due to the differences in grain size and precipitate distribution. Therefore, the observed ductility was measured as an average strain over the gage length. When a tensile load was applied to the joint, failure occurred in the weaker regions of the joint [19], which is the HAZ for the specimen T-FSW-AA5086. As a comparison, the BM-AA5086 and L-FSW-AA5086 fractured in the center of the specimen.

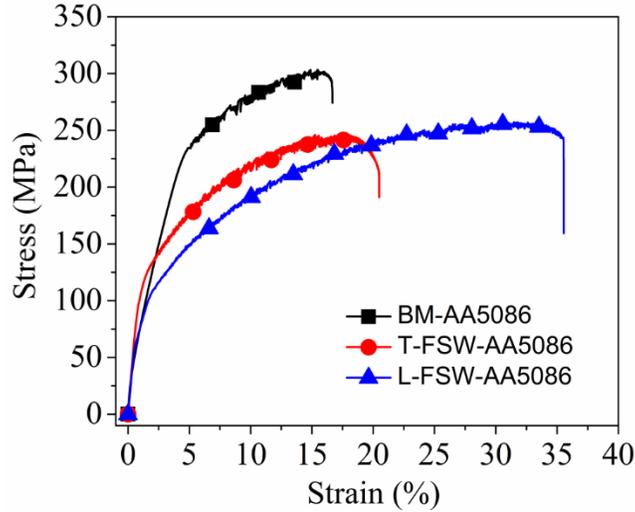

Fig. 4. Tensile stress-strain curves of the two types of FSW specimens as compared to that of the BM.

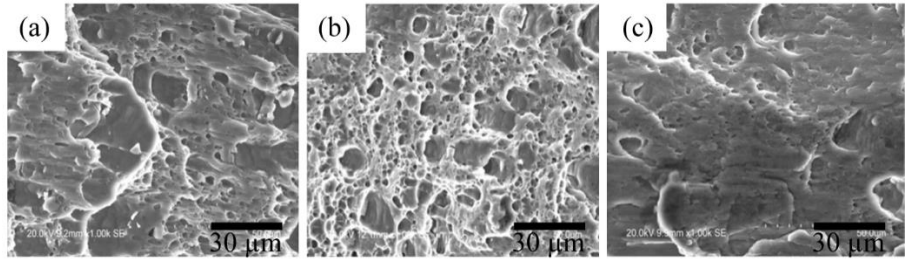

Fig. 5. SEM fractographs of the (a) BM-AA5086, (b) T-FSW-AA5086, and (c) L-FSW-AA5086.

The fractured FSW tensile specimens showed obvious necking/plastic deformation but not the specimen BM-AA5086. The fractographs of the BM and the two FSW specimens (Fig. 5) reveal dimple fracture patterns with teared edges full of micropores. The dimples were of various sizes and shapes. Compared to the specimen BM-AA5086, the dimples in the specimen T-FSW-AA5086 (failure occurred in HAZ) were deeper and the teared edges were thinner. Therefore, the specimen T-FSW-AA5086 exhibited worse mechanical properties than the specimen BM-AA5086. In addition, the specimen T-FSW-AA5086 had much deeper dimples and thinner teared edges than the specimen L-FSW-AA5086 and thus worse mechanical properties.

### 3.3. Electrochemical measurements

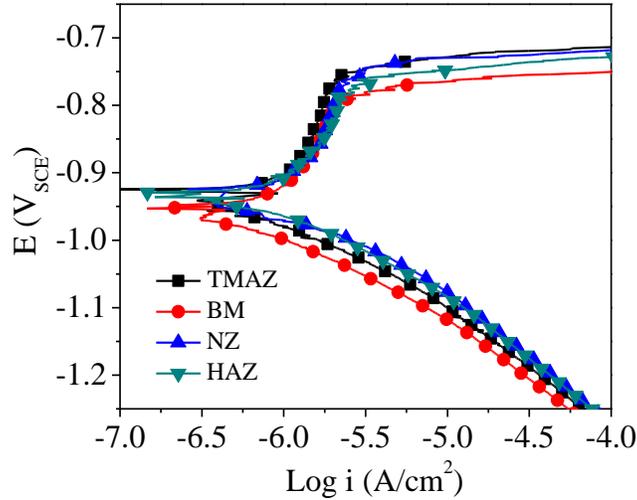

Fig. 6. Polarization curves of different zones in FSW-AA5086 in deaerated 3.15 wt.% NaCl solutions.

Table 1. $E_{corr}$, $I_{corr}$, and $E_{pit}$ values of different zones in FSW-AA5086 in deaerated 3.15 wt.% NaCl solutions.

| Weld | $E_{corr}$ (mV$_{SCE}$) | $E_{pit}$ (mV$_{SCE}$) | $I_{corr}$ (μA/cm$^2$) |
|------|------|------|------|
| HAZ | -929 | -762 | 0.73 |
| TMAZ | -924 | -748 | 1.20 |
| BM | -953 | -767 | 1.70 |
| NZ | -924 | -730 | 1.15 |

Fig. 6 shows typical polarization curves of different zones in FSW-AA5086 in deaerated 3.15 wt.% NaCl solutions. All zones in FSW-AA5086 showed a critical passive region below $E_{pit}$. The current density increased abruptly at potentials greater than $E_{pit}$ due to the breakdown of the passive film. Table 1 summarizes the $E_{corr}$, $I_{corr}$, and $E_{pit}$ values obtained from Fig. 6. Compared to HAZ, TMAZ, and NZ, BM exhibited the lowest $E_{corr}$ and $E_{pit}$ values and the highest $I_{corr}$ value. Therefore, the FSW process improved the corrosion resistance of AA5086 in the weld zones. Notice that HAZ had the lowest $I_{corr}$ value among the four different zones, indicating its highest corrosion resistance. The coarsening of precipitates in the HAZ region seem to be the responsible factors for the improved corrosion resistance [20, 21]. Finally, it is worth to mention that a higher $E_{pit}$ value (e.g., NZ and TMAZ) did not necessarily correspond to a lower corrosion rate, while $I_{corr}$ gave the direct measure of corrosion rate [22].

## 3.4. SEM analysis

The SEM images of FSW-AA5086 samples that were immersed in the three different solutions for 90 days are shown in Figs. 7-9. Generally, the SEM images showed that the amounts of corrosion products on FSW-AA5086 samples immersed in the three solutions decreased from 3.15 wt.% NaCl to ASTM seawater and finally to the 0.5 M Na$_2$SO$_4$ solution. This trend indicates that among the three solutions, 3.15 wt.% NaCl solutions is the most corrosive environment for AA5086 samples while 0.5 M Na$_2$SO$_4$ is the least corrosive environment. In addition, the SEM images showed that, in all three environments, HAZ had much less corrosion products than the other three zones (i.e., NZ, TMAZ, and BM). However, it is difficult to differentiate the other three zones in terms of the amounts of corrosion products on them. The conclusion that HAZ is the most corrosion-resistant zone in FSW-AA5086 is in good agreement with the polarization results as discussed in Section 3.3.

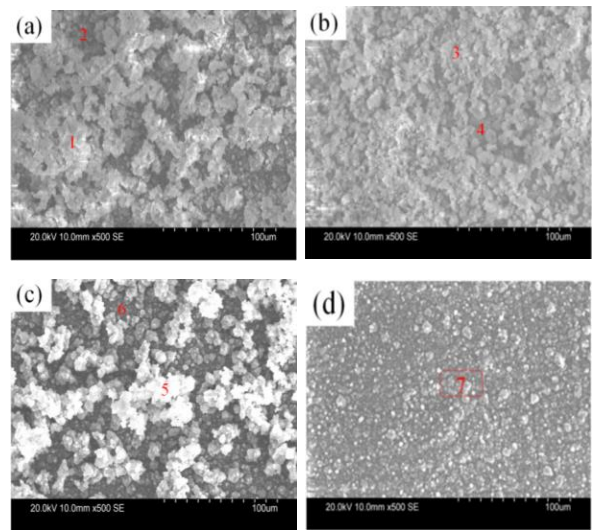

Fig. 7. SEM images of different zones in FSW-AA5086 after 90 days immersion in 3.15 wt.% NaCl: (a) TMAZ, (b) BM, (c) Nugget, and (d) HAZ.

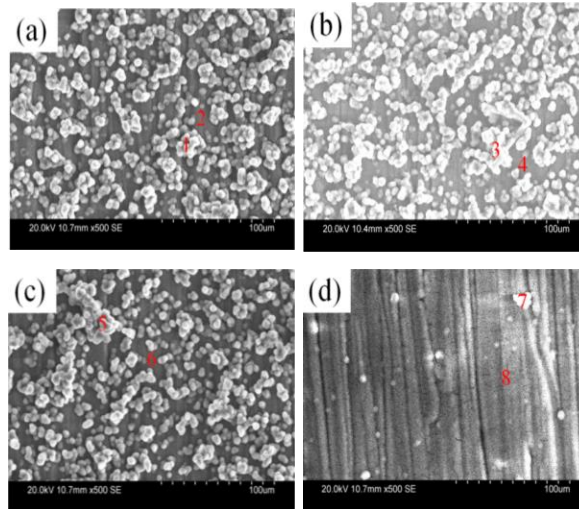

Fig. 8. SEM images of different zones in FSW-AA5086 after 90 days immersion in ASTM seawater: (a) TMAZ, (b) BM, (c) Nugget, and (d) HAZ.

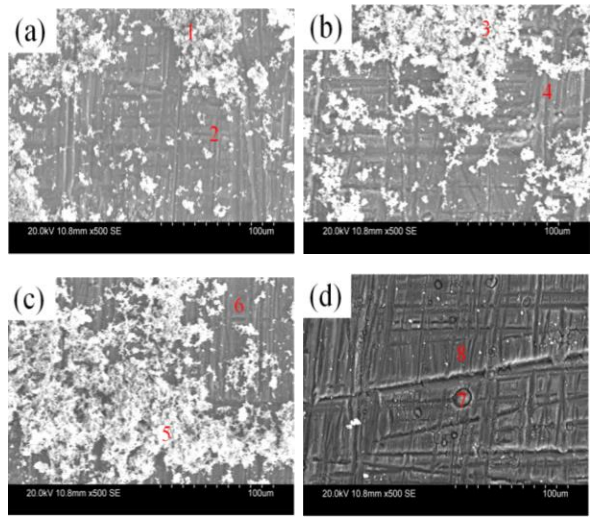

Fig. 9. SEM images of different zones in FSW-AA5086 after 90 days immersion in 0.5 M $Na_2SO_4$: (a) TMAZ, (b) BM, (c) Nugget, and (d) HAZ.

## 3.5. Raman spectroscopic and XRD analysis

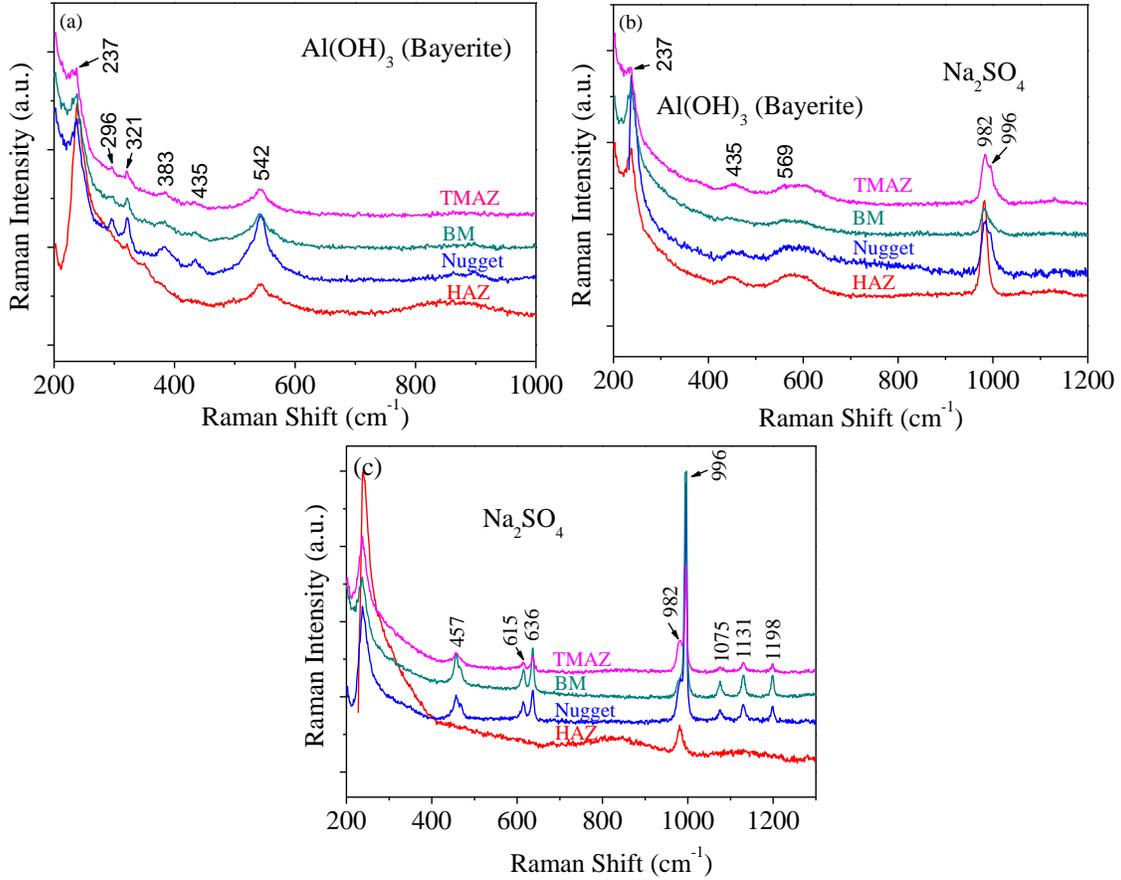

Fig. 10. Raman spectra obtained from the corrosion products on different zones in FSW-AA5086 after 90 days immersion in (a) 3.15 wt.% NaCl, (b) ASTM seawater, and (c) 0.5 M Na₂SO₄ solution.

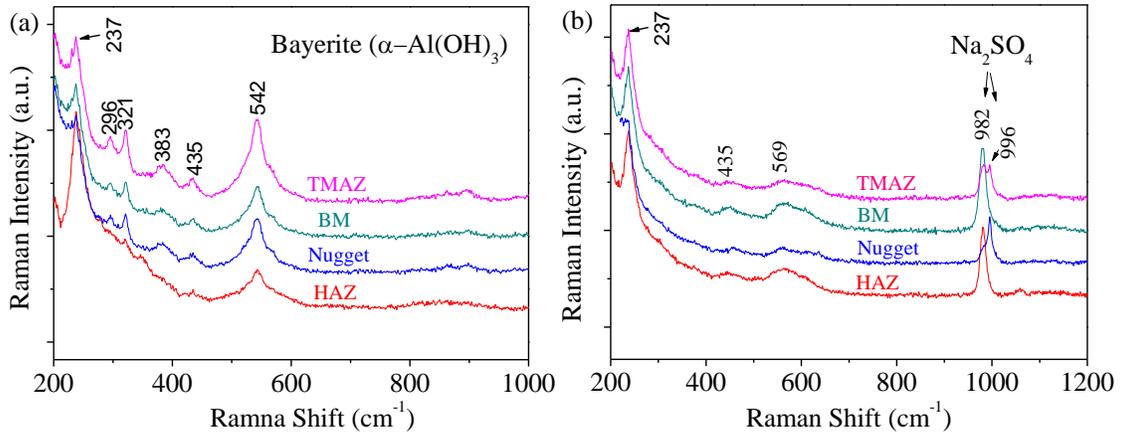

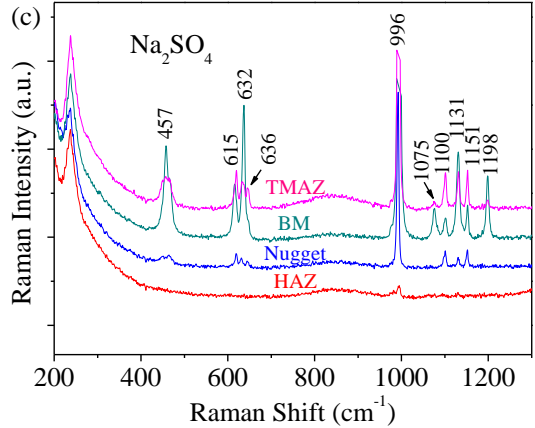

Fig. 11. Raman spectra obtained from the corrosion products on different zones in FSW-AA5086 after 120 days immersion in (a) 3.15 wt.% NaCl, (b) ASTM seawater, and (c) 0.5 M $Na_2SO_4$ solution.

Fig. 10 shows Raman spectra obtained from the corrosion products formed on different zones in FSW-AA5086 after 90 days immersion in 3.15 wt.% NaCl, ASTM seawater, and 0.5 M $Na_2SO_4$ solutions. Bayerite (α-Al(OH)$_3$) with characteristic Raman bands at 237, 296, 321, 383, 435, and 542 cm$^{-1}$ were detected from all zones in FSW-AA5086 immersed in 3.15 wt.% NaCl solution (Fig. 10a) [23]. Notice that the strong Raman band at 237 cm$^{-1}$ might be partially from system noise [24-26]. On the sample immersed in ASTM seawater (Fig. 10b), only weak Raman signal of α-Al(OH)$_3$ at 435 and 569 cm$^{-1}$ was detected. The sharp bands at 982 and 996 cm$^{-1}$ are from $Na_2SO_4$ (rruff.info) in the ASTM sea salt. On the sample immersed in $Na_2SO_4$ solution (Fig. 10c), only Raman signal of $Na_2SO_4$ (457, 615, 636, 996, 1075, and 1131 cm$^{-1}$) was observed and that from Al(OH)$_3$ was not detected.

Fig. 11 shows the Raman spectra obtained from different zones in FSW-AA5086 after 120 days immersion in 3.15 wt.% NaCl, ASTM seawater, and 0.5 M $Na_2SO_4$ solutions. Generally, the Raman spectra in Fig. 11 are similar to those shown in Fig. 10. One apparent difference is that the Raman bands in Fig. 11a are relatively stronger than those in Fig. 10a, indicating that more corrosion products (α-Al(OH)$_3$) formed on AA5086 after longer exposure. The absence of prominent Raman bands of Al(OH)$_3$ on samples immersed in ASTM seawater and 0.5 M Na$_2$SO$_4$ solutions is likely an indication of less corrosion products on these samples as compared to that on the sample immersed in NaCl solution. The Raman spectroscopic analysis agreed well with the SEM results and showed that more corrosion products (i.e., Al(OH)$_3$) formed on FSW-AA5086 samples immersed in 3.15 wt.% NaCl solution than the other two solutions. Another agreement between the Raman analysis and SEM observation is that they all differentiate the HAZ from the other three zones on FSW-AA5086 immersed in 3.15 wt.% NaCl, in terms of the amount of corrosion products. This is evidenced by the two Raman bands at 296 and 321 cm$^{-1}$ for Bayerite having the lowest intensities in the Raman spectra (Fig. 10a and Fig. 11a) from HAZ as compared to the other three zones.

Fig. 12 shows the XRD patterns obtained from FSW-AA5086 samples immersed in the three solutions for 90 days. Notice that XRD patterns were collected from both the upside and downside of the FSW weld zone and the base metal with an area of approximately 1 × 1 in$^2$. Fig. 12a shows strong peaks of α-Al(OH)$_3$ which indicates that a considerable amount of α-Al(OH)$_3$ formed on the samples immersed in NaCl solution. On the contrary, only extremely weak peaks of α-Al(OH)$_3$ were observed in Fig. 12b and c, implying the formation of small amounts of corrosion products. The XRD analysis generally agrees well with SEM and Raman results.

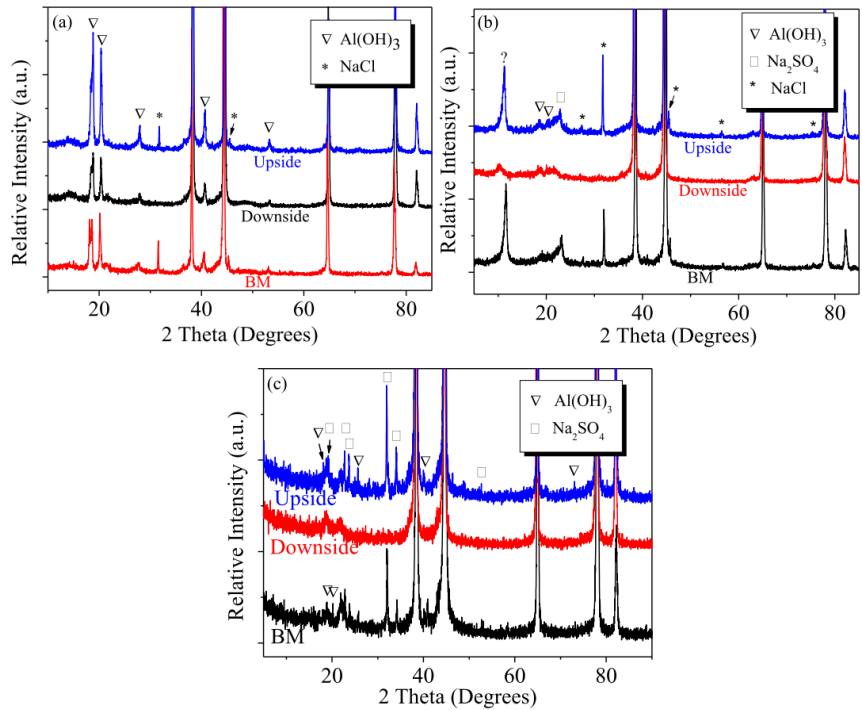

Fig. 12. XRD patterns obtained from FSW-AA5086 samples after 90 days immersion in (a) 3.15 wt.% NaCl, (b) ASTM seawater, and (c) 0.5 M Na$_2$SO$_4$ solution.

## 3.6 Weight loss measurements

The corrosion rates of FSW-AA5086 immersed in the three different solutions for 90 and 120 days are shown in Fig. 13. The corrosion rates of FSW-AA5086 decreased from 90 days to 120 days in three solutions, which might be due to the main corrosion product Al(OH)$_3$ being protective to the substrate. The corrosion rates of FSW-AA5086 from the highest to the lowest for three solutions was as follows: 3.15 wt.% NaCl > ASTM seawater > 0.5 M Na$_2$SO$_4$, which agrees with SEM, Raman, and XRD results.

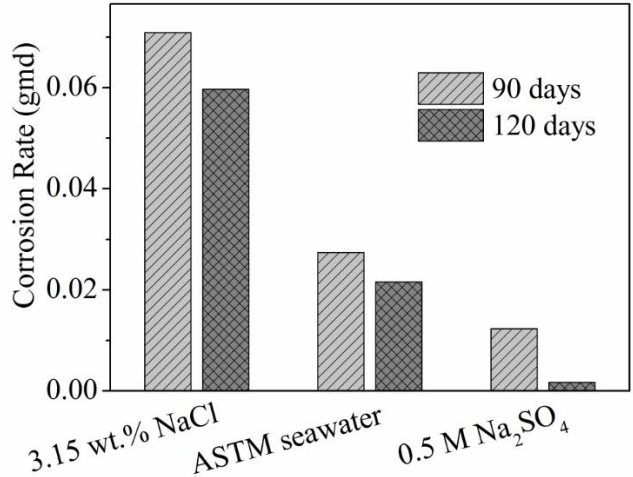

Fig. 13. Corrosion rates of FSW-AA5086 specimens after 90 and 120 days immersion in 3.15 wt.% NaCl, ASTM seawater, and 0.5 M $Na_2SO_4$ solution.

## 4. Conclusions

The corrosion behaviors and mechanical properties of friction stir welded AA5086 were investigated and the difference between different zones was discussed. Based on the results, the following conclusions were reached:

1.  All FSW-AA5086 specimens showed significant decreases in both tensile and yield strengths, but an increase in ductility. The T-FSW-AA5086 tensile specimens failed in the TMAZ/HAZ regions, where the lowest hardness values were detected. The BM and L-FSW-AA5086 failed in the center of the specimen.

2.  In deaerated 3.15 wt.% NaCl solution, all weld zones (i.e., NZ, TMAZ, and HAZ) showed higher $E_{corr}$ and $E_{pit}$ values as compared to the BM, indicating increased corrosion resistance by the FSW process. The minimum value of $I_{corr}$ appeared in the HAZ region.

3.  SEM observation, Raman spectroscopic and XRD characterization of the corrosion products on FSW-AA5086 immersed for 90 and 120 days in three solutions showed that more severe corrosion occurred on samples immersed in 3.15 wt.% NaCl, followed by the ASTM seawater, and then 0.5 M $Na_2SO_4$. SEM and Raman results also showed that HAZ had less corrosion as compared to the other three regions. The main corrosion products was identified as $\alpha$-$Al(OH)_3$ (bayerite) using Raman and XRD analyses.

**Acknowledgements**

The authors are grateful for support of the financial support from the Office of the Under Secretary of Defense for the project entitled **"Correlation of Field and Laboratory Studies on the Corrosion of Various Alloys in a Multitude of Hawaii Micro-Climates"** (U.S. Air Force Academy, Contract no.: FA7000-10-2-0010). The authors are particularly grateful to Mr. Daniel Dunmire, Director, Corrosion Policy and Oversight, Office of the Under Secretary of Defense.

**References**

[1] E.N. W.N. Thomas, JC; Murch, MG;Temple-Smith, P;Dawes, CJ.Friction-stir butt welding, GB Patent No. 9125978.8, International patent application No. PCT/GB92/02203, (1991).

[2] M. Ericsson, R. Sandström, Influence of welding speed on the fatigue of friction stir welds, and comparison with MIG and TIG, International Journal of Fatigue, 25 (2003) 1379-1387.

[3] Ø. Frigaard, Ø. Grong, O. Midling, A process model for friction stir welding of age hardening aluminum alloys, Metallurgical and Materials Transactions A, 32 (2001) 1189-1200.

[4] C. Rhodes, M. Mahoney, W. Bingel, R. Spurling, C. Bampton, Effects of friction stir welding on microstructure of 7075 aluminum, Scripta Materialia, 36 (1997) 69-75.

[5] S. Benavides, Y. Li, L. Murr, D. Brown, J. McClure, Low-temperature friction-stir welding of 2024 aluminum, Scripta Materialia, 41 (1999) 809-815.

[6] K. Ramesh, S. Pradeep, V. Pancholi, Multipass friction-stir processing and its effect on mechanical properties of aluminum alloy 5086, Metallurgical and Materials Transactions A, 43 (2012) 4311-4319.

[7] R. Prado, L. Murr, D. Shindo, K. Soto, Tool wear in the friction-stir welding of aluminum alloy 6061+ 20% $Al_2O_3$: a preliminary study, Scripta Materialia, 45 (2001) 75-80.

[8] Z. Chen, S. Li, L.H. Hihara, Microstructure, mechanical properties and corrosion of friction stir welded 6061 Aluminum Alloy, arXiv preprint arXiv:1511.05507, (2015).

[9] R. Fonda, P. Pao, H. Jones, C. Feng, B. Connolly, A. Davenport, Microstructure, mechanical properties, and corrosion of friction stir welded Al 5456, Materials Science and Engineering: A, 519 (2009) 1-8.

[10] G. Çam, S. Güçlüer, A. Çakan, H. Serindag, Mechanical properties of friction stir butt-welded Al-5086 H32 plate, Materialwissenschaft und Werkstofftechnik, 40 (2009) 638-642.

[11] G. Çam, S. Güçlüer, A. Çakan, H. Serindağ, Mechanical properties of friction stir butt-welded Al-5086 H32 plate, Journal of Achievements in Materials and Manufacturing Engineering, 30 (2008).

[12] E. Taban, E. Kaluc, Comparison between microstructure characteristics and joint performance of 5086-H32 aluminium alloy welded by MIG, TIG and friction stir welding processes, Kovove Materialy, 45 (2007) 241.

[13] I. Shigematsu, Y.-J. Kwon, K. Suzuki, T. Imai, N. Saito, Joining of 5083 and 6061 aluminum alloys by friction stir welding, Journal of Materials Science Letters, 22 (2003) 353-356.

[14] H.-B. Chen, K. Yan, T. Lin, S.-B. Chen, C.-Y. Jiang, Y. Zhao, The investigation of typical welding defects for 5456 aluminum alloy friction stir welds, Materials Science and Engineering: A, 433 (2006) 64-69.

[15] M. Peel, A. Steuwer, M. Preuss, P. Withers, Microstructure, mechanical properties and residual stresses as a function of welding speed in aluminium AA5083 friction stir welds, Acta Materialia, 51 (2003) 4791-4801.

[16] H. Jin, S. Saimoto, M. Ball, P. Threadgill, Characterisation of microstructure and texture in friction stir welded joints of 5754 and 5182 aluminium alloy sheets, Materials Science and Technology, 17 (2001) 1605-1614.

[17] M. James, D. Hattingh, G. Bradley, Weld tool travel speed effects on fatigue life of friction stir welds in 5083 aluminium, International Journal of Fatigue, 25 (2003) 1389-1398.

[18] D. Jeong, U. Erb, K. Aust, G. Palumbo, The relationship between hardness and abrasive wear resistance of electrodeposited nanocrystalline Ni–P coatings, Scripta Materialia, 48 (2003) 1067-1072.

[19] Z.L. Hu, X.S. Wang, S.J. Yuan, Quantitative investigation of the tensile plastic deformation characteristic and microstructure for friction stir welded 2024 aluminum alloy, Materials Characterization, 73 (2012) 114-123.

[20] C.S. Paglia, R.G. Buchheit, Microstructure, microchemistry and environmental cracking susceptibility of friction stir welded 2219-T87, Materials Science and Engineering: A, 429 (2006) 107-114.

[21] P.B. Srinivasan, K.S. Arora, W. Dietzel, S. Pandey, M.K. Schaper, Characterisation of microstructure, mechanical properties and corrosion behaviour of an AA2219 friction stir weldment, Journal of Alloys and Compounds, 492 (2010) 631-637.

[22] K. Surekha, B.S. Murty, K.P. Rao, Microstructural characterization and corrosion behavior of multipass friction stir processed AA2219 aluminium alloy, Surface and Coatings Technology, 202 (2008) 4057-4068.

[23] H.D. Ruan, R.L. Frost, J.T. Kloprogge, Comparison of Raman spectra in characterizing gibbsite, bayerite, diaspore and boehmite, Journal of Raman Spectroscopy, 32 (2001) 745-750.

[24] S. Li, Marine atmospheric corrosion initiation and corrosion products characterization, Mechanical Engineering, (2010) 205.

[25] S. Li, L. Hihara, In situ Raman spectroscopic study of NaCl particle-induced marine atmospheric corrosion of Carbon Steel, Journal of The Electrochemical Society, 159 (2012) C147-C154.

[26] S. Li, L. Hihara, In situ Raman spectroscopic identification of rust formation in Evans' droplet experiments, Electrochemistry Communications, 18 (2012) 48-50.